\documentclass[10pt,journal,compsoc]{IEEEtran}

\ifCLASSOPTIONcompsoc
  \usepackage[nocompress]{cite}
\else
  \usepackage{cite}
  \usepackage{float}
  \usepackage{natbib}
\fi

\usepackage{booktabs}
\usepackage{makecell}
\usepackage[table]{xcolor}
\usepackage{color}
\usepackage{amsmath}
\usepackage{amsthm}
\usepackage{amssymb}
\usepackage{MnSymbol}
\usepackage{multirow}
\usepackage{algorithm}
\usepackage{algpseudocode}
\usepackage{graphicx}
\usepackage{subcaption}
\usepackage{dblfloatfix}
\usepackage{array}
\usepackage{url}
\usepackage{hyperref}

\definecolor{ccColor}{rgb}{0.788,0.278,0.216}

\begin{document}

\title{Unified AI Gateway: A Framework for Joint Model Routing and KV Cache Management}

\author{
    Jiaxun Lu, 
    Xiang Zhang,
    Yunfeng Shao
    \thanks{Corresponding author: Yunfeng Shao.}
    \thanks{Jiaxun Lu, Xiang Zhang and Yunfeng Shao are with Huawei, Shenzhen, China (e-mails: lujiaxun@huawei.com; zhangxiang239@huawei.com; shaoyunfeng@huawei.com).}
}

\IEEEtitleabstractindextext{
\begin{abstract}
 Large language model (LLM) inference increasingly spans models that differ in size, capability, price, and provider. This shift creates two costs for developers. One is the integration cost of choosing among and switching between many models. The other is the inference cost of rebuilding a KV cache when it is unavailable or incompatible with the selected model. We define and analyze the Unified AI Gateway as a system setting for an edge-deployed AI traffic hub. It coordinates model routing, KV cache management, and compute placement across end devices, edge resources, and cloud model services. At request time, the gateway jointly selects a target model, an execution site, and a KV cache action under task-quality, latency, cost, and resource constraints. In parallel, background cache-management actions optimize KV cache placement, replication, retrieval, and lifecycle decisions for subsequent requests. We synthesize existing evidence on KV cache reuse, compression, cross-model mapping, distributed storage, and transfer, and discuss the remaining challenges of integrating these capabilities into one system. Across eight typical workload profiles, our workload-level analytical simulation reports TTFT speedups of 1.25$\times$--13.28$\times$ and input-cost benefits of 1.20$\times$--6.16$\times$.
\end{abstract}

}

\maketitle

\IEEEdisplaynontitleabstractindextext
\IEEEpeerreviewmaketitle

\section{Introduction}
\IEEEPARstart{A}{s} the model ecosystem expands, developers increasingly choose among models that differ in size, capability, price, and provider. Consider a long-context coding agent. Its planning, coding, execution, and review steps may favor different models, while conversation history, tool outputs, and repository state recur across turns~\cite{vllmmooncakestore2026}. Existing AI gateways provide unified interfaces and routing functions such as load balancing and failover, but they do not coordinate the KV cache with provider and model selection. When routing switches the target model, the accumulated cache may be unavailable or incompatible, forcing the target model to repeat full-context prefill. This repeated work can outweigh the capability or price advantage of switching models. Model routing and KV cache management must therefore be coordinated at the gateway.

\begin{figure}[t!]
\centering
\includegraphics[width=\columnwidth]{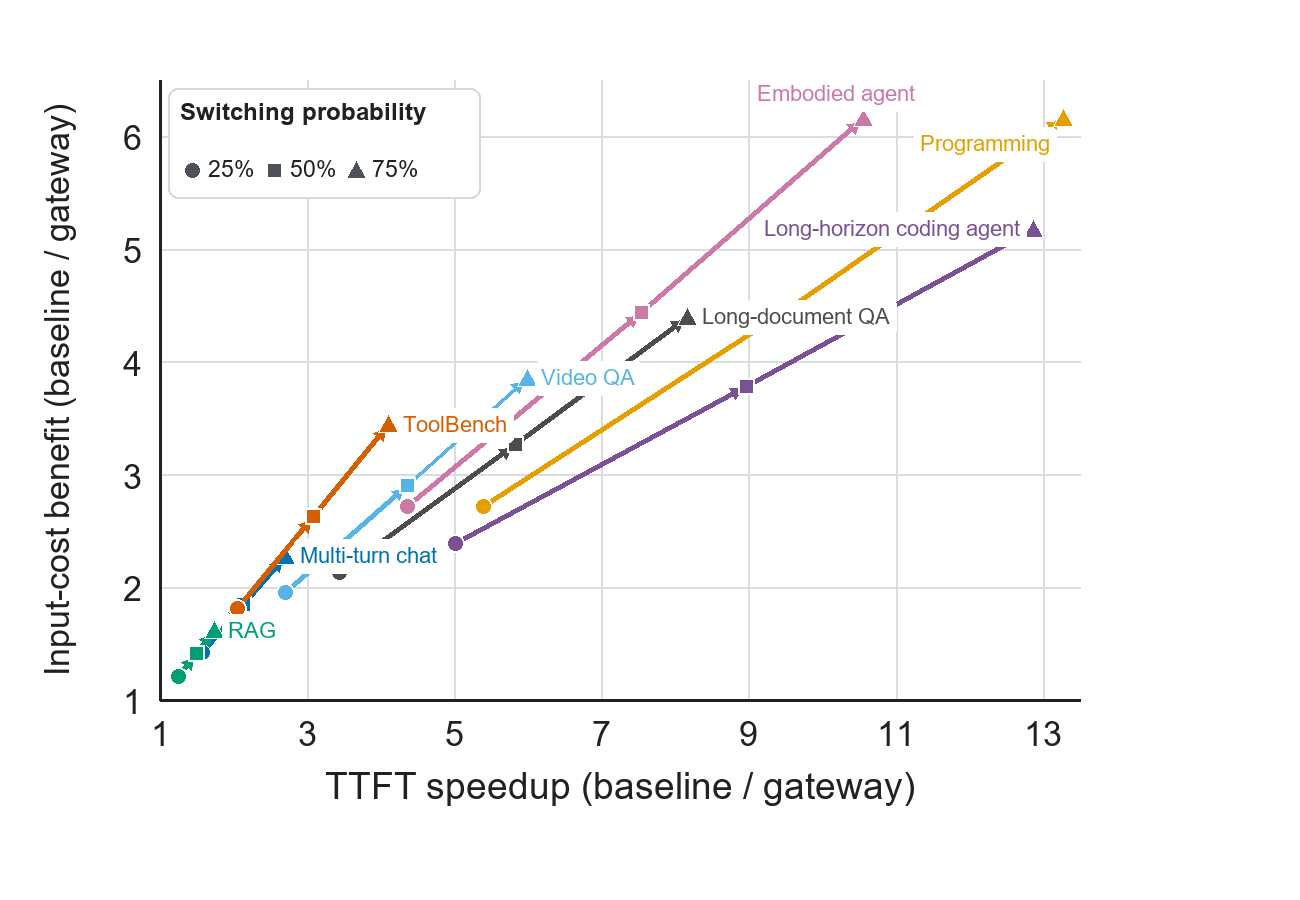}
\caption{Across eight workload profiles, the Unified AI Gateway yields modeled TTFT speedups of 1.25$\times$--13.28$\times$ and input-cost benefits of 1.20$\times$--6.16$\times$. This End--Edge--Cloud setting coordinates model routing, compute placement, and KV cache handling across end devices, edge resources, and cloud model services, so that available caches can be reused, mapped, or transferred instead of fully recomputed. Marker shapes denote model-switching probabilities of 25\%, 50\%, and 75\%, defined as the probability that a request with an available KV cache is routed to a different model. Points farther toward the upper right indicate larger benefits.}
\label{fig:intro_benefit_frontier}
\end{figure}

Current support remains fragmented across gateway services, inference engines, and cache systems. Unified APIs reduce provider-integration overhead, while runtimes such as vLLM and SGLang and systems such as LMCache and Mooncake support runtime reuse, cache lifecycle management, cross-engine sharing, or distributed cache storage~\cite{vllmapc2026,sglangsessioncache2026,anthropicpromptcaching2026,lmcache2025,mooncake2025}. However, these capabilities are not jointly coordinated with provider and model selection across devices and services. A routing decision can therefore discard useful state, increase time to first token (TTFT) and input cost, and create cache-induced model lock-in. To address this issue, this paper proposes the Unified AI Gateway, which couples provider and model routing with KV cache handling across participating runtimes. Figure~\ref{fig:intro_benefit_frontier} previews the resulting benefits. Across the typical workloads and model-switching rates considered in this study, our workload-level simulation reports modeled TTFT speedups of 1.25$\times$--13.28$\times$ and input-cost benefits of 1.20$\times$--6.16$\times$.

\begin{figure*}[t!]
\centering
\includegraphics[width=1.5\columnwidth]{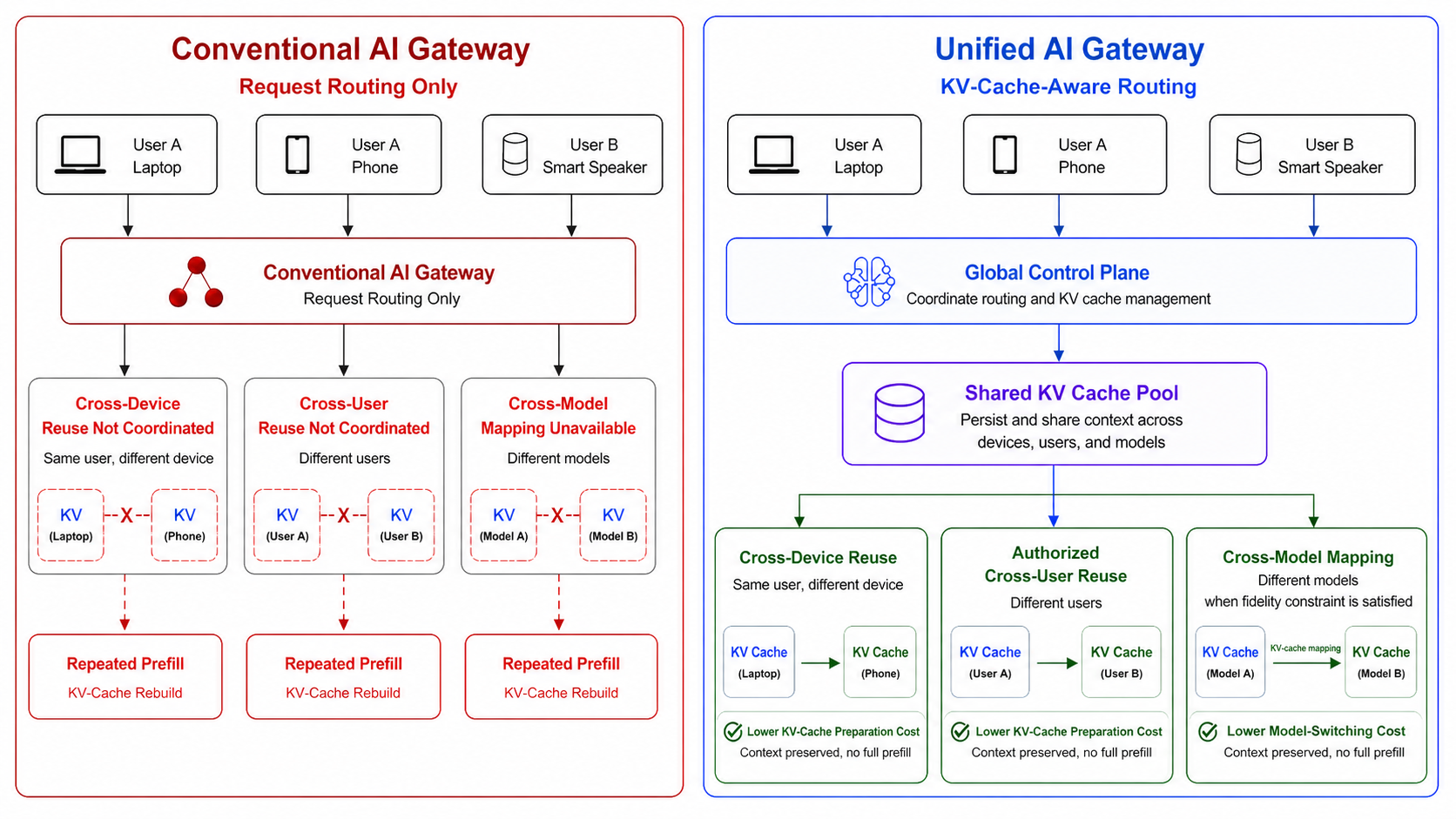}
\caption{Comparison of a Conventional AI Gateway (left) and the Unified AI Gateway (right). The conventional gateway routes requests without coordinating shared KV caches, whereas the Unified AI Gateway coordinates cache reuse and cross-model mapping to avoid target-side full re-prefill when a valid mapping is available.}
\label{fig:motivation}
\end{figure*}

To realize these benefits, the gateway coordinates routing, KV cache storage, prefill, and cache transformation along the user-to-service path. Its placement at the edge supports authorized cache sharing across devices and users, while edge compute can perform local prefill or map KV caches between models subject to fidelity constraints~\cite{kvlink2025,semsharekv2025,droidspeak2025,c2c2026}. This coordination raises three system challenges involving KV cache compatibility and fidelity, the choice between transfer and recomputation, and cache placement, replication, and isolation under capacity and privacy constraints. The Unified AI Gateway therefore considers one joint routing control layer with four KV cache management capability groups, namely cross-granularity reuse, adaptive compression, cross-model mapping, and distributed storage and transfer. Figure~\ref{fig:motivation} contrasts this coordination with the request-routing scope of a conventional AI gateway.

The four capability groups are grounded in active research lines, but their support remains fragmented across inference engines, cache systems, and model-serving services. Mainstream inference engines support exact prefix caching, while recent studies have extended KV-cache reuse to semantic, non-prefix, workflow-aware, and agent-oriented settings~\cite{sglang2024,pagedattention2023,semsharekv2025,kvflow2025,simllm2025,cacheblend2025,sparsex2026,redknot2026,agentkvshift2026}. Distributed systems have also advanced KV-cache storage and transfer~\cite{kvcomm2025,lmcache2025}. The unresolved gateway-level problem is to jointly decide whether to reuse, map, transfer, or recompute a cache while selecting the provider, model, and placement under quality, latency, cost, and resource constraints.

To address these issues, this paper introduces the Unified AI Gateway, which jointly coordinates model routing and KV-cache management across the End--Edge--Cloud continuum. We evaluate its potential benefits through analytical simulation. The contributions are as follows.

\begin{itemize}
    \item \textbf{Unified gateway abstraction and joint control.} We introduce the Unified AI Gateway as an End--Edge--Cloud abstraction that coordinates model routing, KV-cache management, and compute placement. We formalize it through a Global Control Plane that jointly selects models and cache actions under latency, cost, quality, and resource constraints, providing a basis for gateway-wide optimization.
    \item \textbf{Coordination taxonomy and systems gap.} We organize cache reuse, compression, mapping, storage, transfer, prefill, and routing into an end-to-end taxonomy. Our analysis shows that existing methods optimize individual stages but do not coordinate cache handling, model selection, and compute placement across models and system tiers.
    \item \textbf{Quantified benefits and deployment guidance.} Across eight workload profiles, analytical simulation shows modeled TTFT speedups from 1.25$\times$ to 13.28$\times$ and input-cost benefits from 1.20$\times$ to 6.16$\times$. In particular, the long-horizon coding-agent workload achieves TTFT speedups from 5.01$\times$ to 12.87$\times$. Sensitivity analysis shows that future deployments with higher cache hit rates and greater available bandwidth could further broaden the benefit range.
\end{itemize}

The remainder of this paper is organized as follows. Sections~\ref{sec:setting}--\ref{sec:evidence} define the system setting, formulate the control plane, analyze component requirements, and synthesize existing evidence, respectively. Section~\ref{sec:benefit_analysis} evaluates workload-level benefits through analytical simulation, while Section~\ref{sec:limitations} discusses deployment readiness and future directions. Section~\ref{sec:conclusion} concludes the paper.

\section{Preliminaries and System Setting}\label{sec:setting}

This section introduces the preliminaries needed to treat KV caches as gateway-managed resources, defines the Unified AI Gateway setting, and presents the End--Edge--Cloud architecture that instantiates it. The corresponding joint model-routing and KV cache decision space is presented in Section~\ref{sec:control_plane}.

\subsection{Preliminaries}\label{sec:preliminaries}

\textbf{KV cache.} The KV cache stores the key and value tensors of processed tokens so that autoregressive decoding does not recompute the full context at every step~\cite{attention2017}. Its size grows with context length and depends on the model's layers, KV heads, head dimension, and numerical format. This growing footprint makes efficient storage management essential.  Techniques such as grouped-query attention (GQA), multi-head latent attention (MLA), and hybrid attention change the KV cache memory footprint but not the underlying management problem~\cite{qwen32025,deepseekv42026}. PagedAttention organizes the cache into fixed-size blocks to improve memory utilization~\cite{pagedattention2023}. These properties make the KV cache an explicit infrastructure object whose storage, reuse, placement, and transfer affect inference latency and resource use.

\textbf{KV cache reuse.} KV cache reuse shares previously computed attention caches across requests to avoid repeated prefill. Exact prefix caching is the most mature form and is widely supported by inference engines~\cite{sglang2024,pagedattention2023}. Semantic, non-prefix, workflow-level, and cross-model reuse broaden the coverage to similar prompts, interleaved contexts, multi-agent workflows, and model switching~\cite{semsharekv2025,cacheblend2025,kvcomm2025,droidspeak2025,c2c2026}. Greater coverage introduces matching, calibration, correction, and fidelity costs. Accordingly, the gateway treats reuse granularity as a KV cache decision rather than as a fixed cache policy. The detailed evidence is reviewed in Section~\ref{sec:feasibility_reuse}.

\textbf{Prefill--decode disaggregation.} Prefill-decode (PD) disaggregation places the compute-intensive prefill phase and the memory-intensive decode phase on different nodes, enabling independent resource management~\cite{splitwise2024,distserve2024}. It makes the prefill-generated KV cache an explicit network payload that must be stored, transferred, and consumed by a decode node. Systems such as Mooncake and NVIDIA Dynamo demonstrate the infrastructure relevance of this design, while HACK highlights the transfer bottleneck created by long contexts~\cite{mooncake2025,priceofanarchy2026,hack2025}. End--Edge--Cloud deployment extends this workflow to edge prefill and terminal or cloud decoding, which is the setting analyzed by the unified gateway.

\subsection{Unified AI Gateway Definition}\label{sec:gateway_setting}

We define the Unified AI Gateway as an edge-oriented system setting in which request routing and KV cache management are coordinated on the same request path. The gateway connects end devices, edge resources, distributed cache storage, prefill services, and a pool of model-serving clusters. It manages model-specific KV caches together with metadata describing model version, token span, cache layout, location, ownership, freshness, and provenance. For each request, the gateway may select a model, an execution site, and one KV cache action, including direct reuse, cross-model mapping, compression, transfer, edge or remote prefill, and target-side re-prefill. The setting therefore covers both the architecture in which the edge manages KV caches and the deployment constraints that determine whether a cache action is feasible. The End--Edge--Cloud architecture is presented below, while the corresponding decision space is formalized in Section~\ref{sec:control_plane}.

\subsection{End--Edge--Cloud Architecture}\label{sec:end_edge_cloud}

\begin{figure}[t!]
\centering
\includegraphics[width=\columnwidth]{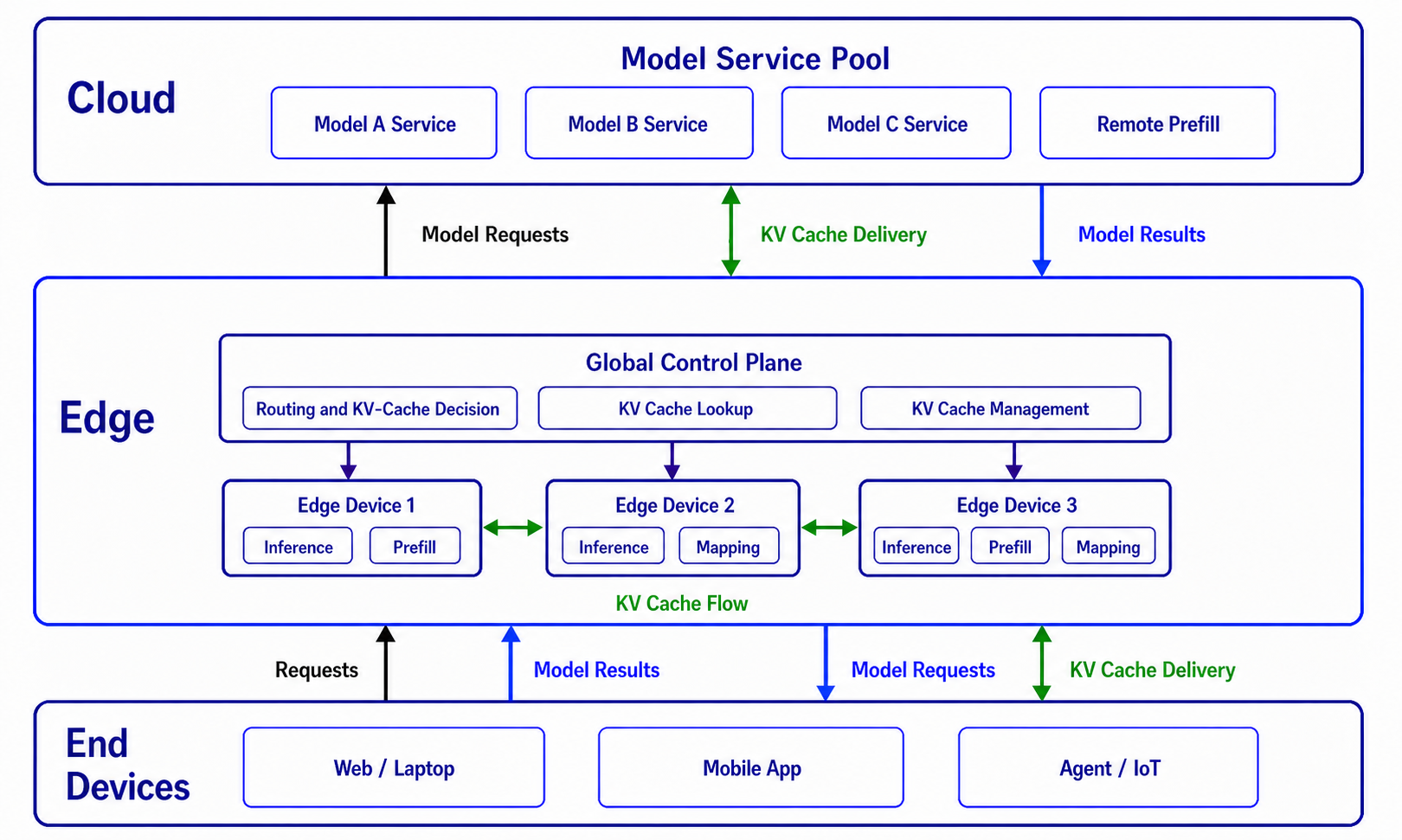}
\caption{High-level overview of the End--Edge--Cloud architecture that constitutes the Unified AI Gateway. The global control plane coordinates model routing, cache lookup, and KV cache management across end devices, distributed edge devices, and cloud services, while edge devices provide inference, prefill, and cross-model mapping. Requests, results, and KV caches can flow bidirectionally among the participating layers. Detailed cache actions are shown in Figure~\ref{fig:decision}.}
\label{fig:architecture}
\end{figure}

The Unified AI Gateway is realized as an End--Edge--Cloud three-layer collaborative architecture (Figure~\ref{fig:architecture}). In this paper, the Unified AI Gateway denotes the complete system setting. The Global Control Plane is its control core, while end devices and cloud model services are participating endpoints. The edge layer hosts the Global Control Plane and distributed edge devices. Its core design is to place KV cache management on the request data path. The control plane decides not only which model to route to, but also how the associated cache is stored, reused, transformed, and transferred.

\textbf{The end side} consists of terminal devices (web applications, mobile apps, Internet of Things (IoT) devices, etc.) that originate inference requests and consume generated results. Their submitted requests contain a user identifier and a context sequence (prompt, dialogue history, etc.). Under the End--Edge--Cloud PD-disaggregated form, terminals may also perform decode computation, in which case the edge gateway delivers the prefill-generated KV cache to them (Section~\ref{sec:edge_prefill}). When terminal-side inference is selected, the gateway can forward a model request and the required KV cache to the terminal-side service, then receive the generated result through the same edge path.

\textbf{The edge side} hosts the global control plane and the concrete edge devices that execute selected cache actions. The global control plane provides joint routing and cache-action decisions, cache lookup, and KV cache management, while the edge devices provide inference, prefill, and cross-model mapping. These functions share request and cache metadata. Routing can therefore account for cache availability and preparation cost, while placement can reflect routing priorities, locality, capacity, and access policy. Here, KV cache management refers to control-plane coordination of cache lookup, placement, transfer, and lifecycle actions. The underlying storage and migration mechanisms can be provided by distributed cache systems such as Mooncake and LMCache~\cite{mooncake2025,lmcache2025}, while the Unified AI Gateway determines when and where these capabilities should be used. The detailed cache-handling workflow is described in Section~\ref{sec:control_plane}, and its technical requirements are analyzed in Section~\ref{sec:component_requirements}. Edge devices can host prefill computation (Section~\ref{sec:edge_prefill}), allowing the edge to participate directly in cache generation. With suitable model weights and runtime support, they can additionally host model inference services and serve requests locally. The control plane stages and transfers the resulting cache to the selected cloud-side or terminal-side decode service when the serving computation is placed remotely.

\textbf{The cloud side} comprises the inference service clusters of model providers (including commercial APIs and self-hosted model services), receiving requests routed by the edge gateway and returning generated results. The gateway can also instruct the cloud side to perform prefill directly and return the KV cache to the edge. Depending on the selected placement, the cloud, edge, or end side can provide the serving or decode service.

The workflow in Figure~\ref{fig:architecture} follows the request path from the end to the edge and then to the selected serving destination. The gateway authenticates the request, extracts its context and service requirements, and uses cache metadata, resource status, and network conditions to filter feasible actions and estimate their end-to-end cost. A valid KV cache is reused directly or mapped to the target model. Otherwise, the gateway performs edge prefill or schedules prefill remotely, then compresses and delivers the resulting cache to the selected decode node, which may be hosted in the edge, cloud, or end side.

The gateway uses cache availability, resource status, network conditions, authorization, and task-level quality feedback to update future routing and placement policies.

\section{Global Control Plane and KV-Cache Workflow}\label{sec:control_plane}

This section presents the Global Control Plane and its joint model-routing, compute-placement, and KV cache decision space. It then traces how the control plane coordinates cache preparation and delivery across the gateway components.

\subsection{Joint Routing, Compute Placement, and KV Cache Management}\label{sec:joint_decision}

The Global Control Plane makes coupled decisions at two timescales. At request time, it chooses the serving model, its execution site, and how the required KV cache should be obtained. The selected model determines the target cache format, while the execution site determines where serving or decode computation runs. Cache availability, mapping cost, prefill resources, and network conditions affect the cost of this model-site pair. Across the request stream, KV cache management continuously optimizes the shared cache pool across edge devices. It determines where caches are stored, whether frequently requested caches should be replicated near likely request sources or decode destinations, and when caches should be migrated or evicted under storage, bandwidth, freshness, authorization, and consistency constraints. These background actions shape the caches available to later requests and can reduce their preparation cost.

\begin{figure*}[t!]
\centering
\includegraphics[width=0.8\textwidth]{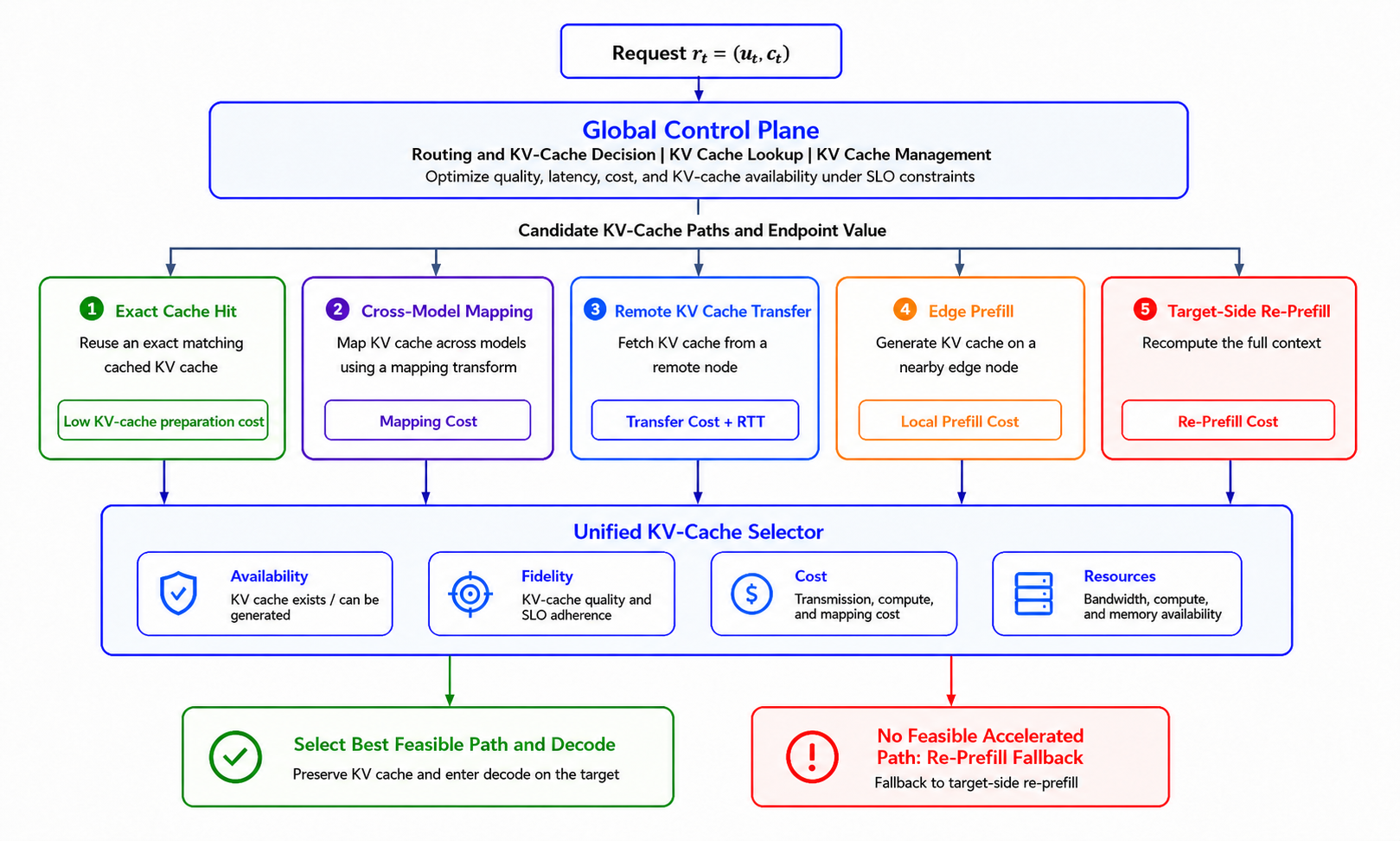}
\caption{Global control-plane workflow for joint model routing, compute placement, and KV cache action selection. The control plane compares cache hit, cross-model mapping, edge prefill, remote cache transfer, and target-side re-prefill according to end-to-end cost, fidelity, and resource constraints.}
\label{fig:decision}
\end{figure*}

To make this joint decision, the control plane must first determine which model, execution-site, and KV cache paths are feasible and then compare their end-to-end value. For each candidate model and site, Figure~\ref{fig:decision} presents five cache paths covering exact cache hits, cross-model mapping, remote cache transfer, edge prefill, and target-side re-prefill. These paths are evaluated along four dimensions shown in the Unified Cache Selector. (i)~\emph{Availability} asks whether the required cache exists, can be looked up, or can be generated under the current placement and authorization policy. (ii)~\emph{Fidelity} covers cross-model cache compatibility, mapping or compression error, and adherence to the quality and service-level objective (SLO) constraints. (iii)~\emph{Cost} includes cache lookup, transmission delay and round-trip time (RTT), compression and decompression, mapping, prefill or recomputation, memory copies, and cache restoration. (iv)~\emph{Resources} include link bandwidth, end, edge, and cloud compute, memory and storage capacity, and queueing conditions. Cache lookup and management affect every candidate path, while request-stream decisions determine the feasible action set.

Let the model set be $\mathcal{M}=\{M_1,\dots,M_k\}$, where $k$ is the number of candidate models, and let request $r_t=(u_t,c_t)$ arrive at time $t$. Here, $u_t$ denotes the user or tenant identity and $c_t$ denotes the request context. Let $\mathcal{D}_t$ be the candidate execution sites across the end, edge, and cloud, and let $\mathcal{D}_t(j)\subseteq\mathcal{D}_t$ contain the sites that can execute model $M_j$ or its decode phase. Prefill of context $c_t$ on model $M_i$ produces KV cache $K_i(c_t)$, whose size grows approximately linearly with context length~\cite{pagedattention2023}. For a coding-agent request, for example, $c_t$ can include the accumulated conversation, tool outputs, and repository state that recur across turns. A model switch then requires the control plane to select an execution site and decide whether the source cache can be reused directly, mapped to the target model, transferred, or replaced by target-side re-prefill.

At the request-stream level, let $\mathcal{E}$ denote the set of edge devices and let $P_t(K)\subseteq\mathcal{E}$ denote the devices that store or stage cache object $K$ at time $t$. KV cache management updates $P_t(K)$ according to request frequency, locality, capacity, freshness, authorization, and transfer cost, thereby allowing frequently requested caches to be replicated near likely request sources or decode destinations. In the per-request formulation below, $P_t$ is the current cache placement produced by these background management actions. It affects the feasible actions and their costs but is not itself selected by the instantaneous joint decision. For a target model $M_j$ and execution site $e\in\mathcal{D}_t(j)$, let $\mathcal{A}_t(j,e;P_t)$ be the feasible KV cache actions given the current cache placement, cache contents, compute resources at $e$, network conditions, authorization policy, and requirements for keeping users' data separate. Thus, $e$ captures request-time compute placement, while $P_t$ captures the cache placement available when the request arrives. For cross-model mapping, the transformed cache is written as $\hat{K}_j(c_t)=f_{i\to j}(K_i(c_t))$.

Define the feasible tuple set as $\mathcal{F}_t(P_t)=\{(j,e,a):M_j\in\mathcal{M},\ e\in\mathcal{D}_t(j),\ a\in\mathcal{A}_t(j,e;P_t)\}$. For each $(j,e,a)\in\mathcal{F}_t(P_t)$, let $L_{tjea}$ denote the cache-preparation and delivery latency, $C_{tjea}$ the corresponding compute and transfer cost, and $q_j(r_t)$ the standalone quality of model $M_j$. If $d_{tjea}$ denotes the task-level quality loss relative to target-side re-prefill on the selected model, the resulting quality is $Q_{tjea}=q_j(r_t)-d_{tjea}$. The latency $L_{tjea}$ includes cache lookup, generation or full-context recomputation, mapping, compression and decompression, transfer to $e$, memory copies, restoration, and queueing whenever these operations are used. The cost $C_{tjea}$ accounts for the associated compute, memory, storage, and network resources.

When KV cache transfer is used, a simplified latency model is
\begin{equation}
T_{\mathrm{transfer}}\approx
\frac{S^{\mathrm{compressed}}_{tjea}}{B}
+T_{\mathrm{RTT}}+T_{\mathrm{queue}}+T_{\mathrm{copy}}+T_{\mathrm{restore}},
\label{eq:cache_transfer_latency}
\end{equation}
where $S^{\mathrm{compressed}}_{tjea}$ is the transmitted KV cache after compression, $B$ is the available bandwidth, $T_{\mathrm{RTT}}$ is round-trip communication delay, $T_{\mathrm{queue}}$ is queueing delay, $T_{\mathrm{copy}}$ is memory-copy overhead, and $T_{\mathrm{restore}}$ is decode-side cache restoration and initialization overhead. We use $T_{\mathrm{cache}}^{P_t\to e}(c_t)$ for the complete delivery path from the current cache placement to execution site $e$. Cross-model transfer also incurs mapping latency $T_{\mathrm{map}}^{i\to j}(c_t)$, which is zero for direct same-model transfer.

The control plane then selects a model, an execution site, and a KV cache action jointly, shown as follows
\begin{equation}
\begin{aligned}
(j_t^*,e_t^*,a_t^*)=\arg&\min_{(j,e,a)\in\mathcal{F}_t(P_t)}
\quad \lambda_L \widetilde{L}_{tjea}+\lambda_C \widetilde{C}_{tjea} \\
&\text{s.t. } L_{tjea}\leq\overline{L}_t,\qquad Q_{tjea}\geq\overline{Q}_t.
\end{aligned}
\label{eq:joint_routing_cache}
\end{equation}
Here, $\lambda_L,\lambda_C\geq 0$ are user-configurable weights for latency and cost, $\widetilde{L}_{tjea}$ and $\widetilde{C}_{tjea}$ are normalized latency and cost values, and $\overline{L}_t$ and $\overline{Q}_t$ are the request-level latency and quality requirements. Using reference ranges $[L_{\min},L_{\max}]$ and $[C_{\min},C_{\max}]$, we define $\widetilde{L}=(L-L_{\min})/(L_{\max}-L_{\min})$ and $\widetilde{C}=(C-C_{\min})/(C_{\max}-C_{\min})$. These reference ranges can be estimated from feasible tuples observed for the current request class and planning horizon, then updated as workload and link conditions change. At the request-stream level, edge storage capacity, link bandwidth, cache consistency, and privacy constraints further restrict the feasible action set and its cost. The joint objective in Eq.~\eqref{eq:joint_routing_cache} captures the central Global Control Plane problem. Model routing, compute placement, and KV cache preparation must be optimized together.

\subsection{KV Cache Reuse and Cross-Model Mapping}\label{sec:reuse_translation}

KV cache reuse avoids repeated prefill by sharing computed attention caches across requests. It covers exact prefix reuse, semantic or non-prefix reuse, and workflow-level reuse across multi-turn, multi-agent, or authorized cross-user requests. When a request switches from source model $M_i$ to target model $M_j$, the gateway may transform the source cache, $\hat{K}_j(c)=f_{i\to j}(K_i(c))$, and allow the target model to decode without recomputing the full context. Cross-model mapping broadens model-switching flexibility but introduces transformation cost and a fidelity constraint.

\subsection{Edge-Side Prefill}\label{sec:edge_prefill}

Edge devices may execute prefill within the PD-disaggregated workflow, generating a KV cache near the terminal and delivering it to an edge, cloud, or terminal-side decode service. The gateway uses local execution when the required model and resources are available. Otherwise, it schedules prefill remotely and forwards the returned cache. Direct use requires compatible model and runtime semantics, whereas incompatible caches require mapping or target-side re-prefill. A prefill-generated cache can enter the distributed cache pool for later reuse, mapping, compression, or transfer.

\subsection{KV Cache Transfer, Compression, and Retrieval}\label{sec:transfer_compression}

Transfer, compression, and distributed retrieval move reusable KV caches to the selected decode destination. The gateway estimates delivery latency using Eq.~\eqref{eq:cache_transfer_latency} and compares it with target-side recomputation. Compression reduces transfer volume, while retrieval and placement determine whether the cache is available within latency, fidelity, and authorization constraints. The cache pool may span edge devices, cloud services, and authorized tenant groups, with placement and replication guided by locality, demand, capacity, freshness, and transfer cost. Transfer scheduling must also account for memory layout and inference-engine overhead.

\section{Component Responsibilities and Requirements}\label{sec:component_requirements}

This section maps the request workflow to gateway-specific responsibilities and technical requirements. It covers joint routing, KV cache reuse and cross-model mapping, compression and transfer, distributed cache management, and edge prefill, subject to security requirements that keep users' data separate. It then examines the coupling among routing, transfer scheduling, cache placement, and inference-engine memory management.

\subsection{Joint Model and KV-Cache Routing}\label{sec:req_routing}

The routing control component is responsible for producing the joint model, execution-site, and KV cache decision formalized in Section~\ref{sec:joint_decision}. Its output includes the target model, the site that executes serving or decode, and the cache action used to prepare that model's context. The policy must combine task quality, provider cost, latency objectives, access policy, cache locality, model compatibility, and the fidelity constraint of mapped or compressed caches. It must also react to online feedback. A cache lookup can fail, a mapper may violate the quality threshold, or a remote cache transfer may become slower than target-side re-prefill. Existing model-routing studies are reviewed in Section~\ref{sec:feasibility_routing}. The gateway-specific requirement is to make routing and compute placement cache-aware rather than treating KV cache preparation as a downstream implementation detail.

\subsection{Cross-Granularity KV Cache Reuse}\label{sec:req_cache_reuse}

The reuse manager determines whether an existing KV cache can replace full prefill across multi-model, multi-session, and multi-agent requests. Its gateway-specific requirements are (i)~\emph{scaled exact prefix matching} for high-concurrency lookup~\cite{sglang2024}, (ii)~\emph{semantic-similarity matching} across lexically different prompts~\cite{semsharekv2025,simllm2025}, (iii)~\emph{non-prefix segment reuse} with positional and cross-segment interaction handling~\cite{cacheblend2025,sparsex2026}, (iv)~\emph{multi-agent adaptation} for cache offsets and memory-position drift~\cite{redknot2026,agentkvshift2026}, and (v)~\emph{dynamic heat management} for prioritization and eviction. Hybrid-attention models additionally require layer-aware cache indexing~\cite{deepseekv42026}.

\subsection{KV Cache Compression and Transfer}\label{sec:req_size}

The compression and transfer manager makes reusable KV caches deliverable under heterogeneous edge bandwidth and storage constraints~\cite{cachegen2024}. Given cache structure, bandwidth, RTT, and storage headroom, it selects among low-bit quantization~\cite{rotatekv2025,mpoq2025}, layer-aware mixed compression~\cite{tailorkv2025}, and low-rank methods~\cite{starkv2026}. The gateway balances transfer latency, storage occupancy, fidelity, and computation overhead, while online controllers adapt the decision as link and workload conditions change~\cite{kvserve2026}.

\subsection{Cross-Model KV Cache Mapping}\label{sec:req_cache_mapping}

The cross-model cache translator preserves usable context when routing changes between models. Because models differ in layers, attention heads, hidden dimensions, and positional-encoding schemes, their KV caches are not directly interchangeable~\cite{c2c2026}. The translator therefore needs mapping mechanisms that preserve context continuity while satisfying latency and fidelity requirements.

The main requirements are (i)~\emph{same-architecture mapping} for models with identical architecture but different parameters~\cite{droidspeak2025}, (ii)~\emph{cross-architecture mapping} for heterogeneous model pairs~\cite{c2c2026,mot2026}, and (iii)~\emph{zero-shot or few-shot mapping} because training a dedicated mapper for every model pair is impractical.

\subsection{Distributed KV Cache Management}\label{sec:req_placement}

The distributed cache store makes KV caches discoverable, authorized, and close to the selected decode destination across regions and operators. Its main requirements are (i)~\emph{placement} based on user geography, access patterns, and model locations, (ii)~\emph{efficient retrieval}, (iii)~\emph{transfer scheduling} under bandwidth constraints~\cite{hack2025,lmcache2025}, and (iv)~\emph{consistency maintenance} during model updates and cache invalidation.

\subsection{Control-Plane Coordination}\label{sec:req_coupling}

The gateway control plane is responsible for coordinating decisions that appear modular but share the same bottlenecks. The four component responsibilities above all depend on efficient KV cache transfer. Transfer is itself coupled with the inference engine's memory management. Inference engines such as PagedAttention manage KV caches in fixed token blocks whose physical locations are often non-contiguous. Direct transfers over remote direct memory access (RDMA) or sockets therefore require many scatter-gather direct memory access (DMA) operations and can reduce effective Peripheral Component Interconnect Express (PCIe) bandwidth. Compacting the blocks first introduces extra copies and consumes memory bandwidth that is already scarce during decoding~\cite{pagedattention2023}. Shared wide-area links create another coupling. Large cache objects can occupy the link for a long time and block smaller requests, worsening tail latency. KVServe jointly schedules compression strategies and bandwidth conditions, illustrating the need for system-level coupling~\cite{kvserve2026}.

The gateway therefore needs page-aware, copy-efficient transfer protocols aligned with the engine's page-table structure~\cite{pagedattention2023}, together with scheduling across transfer, storage, and inference. These decisions must account for cross-domain bandwidth and RTT, because long-context requests can generate several gigabytes of cache data.

\section{Evidence Synthesis and System Progress}\label{sec:evidence}

Section~\ref{sec:control_plane} identifies the gateway responsibilities and their cross-cutting coupling constraints. The following subsections organize the existing KV cache evidence by capability and summarize performance, cost, flexibility, and system-coordination benefits. Section~\ref{sec:benefit_analysis} then evaluates how these benefits vary across workload profiles and operating conditions.

\subsection{KV Cache Reuse}\label{sec:feasibility_reuse}

KV cache reuse balances coverage against matching and correction cost. Existing work mainly covers exact prefixes, arbitrary segments, semantic similarity, and multi-agent contexts. Table~\ref{tab:reuse_granularity} organizes these reuse levels before the detailed discussion.

Because the underlying studies use different models, hardware platforms, workloads, and evaluation metrics, Table~\ref{tab:reuse_granularity} presents a taxonomy of reported outcomes rather than a direct cross-row comparison. The rows move from exact matching to increasingly flexible reuse. Potential coverage and reuse benefit generally increase, but end-to-end gains are not monotonic because matching, calibration, and correction costs also increase.

\begin{table*}[t!]
\centering
\caption{Taxonomy of KV cache reuse granularities and representative evidence from heterogeneous workloads.}
\label{tab:reuse_granularity}
\small
\rowcolors{2}{gray!8}{white}
\begin{tabular}{@{}p{0.16\textwidth} p{0.18\textwidth} p{0.18\textwidth} p{0.18\textwidth} p{0.18\textwidth}@{}}
\toprule
Reuse level & Reusable object & Coverage and flexibility & Additional processing & Representative benefit \\
\midrule
Exact prefix & Identical prompt prefix & Lowest coverage and highest matching precision & Hash lookup and cache eviction & 2--6.4$\times$ throughput improvement; 52.4--74.1\% hit rate~\cite{pagedattention2023,sglang2024} \\
Non-prefix segment & Contiguous segment at an arbitrary position, including shifted agent contexts & Broader coverage across interleaved and multi-agent contexts & Boundary recomputation, sparse correction, and offset calibration & 2.2--3.3$\times$ lower TTFT; 2.8--5$\times$ higher throughput; over 70\% reuse and up to 7.8$\times$ prefill speedup in multi-agent workloads~\cite{cacheblend2025,sparsex2026,kvcomm2025} \\
Semantic similarity & Lexically different but similar content & Higher flexibility across prompts and tasks & Embedding matching and positional calibration & 6.25$\times$ speedup; 42\% memory savings~\cite{semsharekv2025} \\
\bottomrule
\end{tabular}
\end{table*}

\emph{Exact prefix reuse} requires the reused content to appear at the prompt prefix and to match exactly. vLLM combined paged attention with block hashing and improved throughput by 2--4$\times$ at the same latency~\cite{pagedattention2023}. SGLang used a hierarchical radix tree and reported up to 6.4$\times$ higher throughput, with production hit rates of 52.4\% on LLaVA-Next-34B and 74.1\% on Vicuna-33B~\cite{sglang2024}. KVFlow added workflow-aware eviction and reported up to 1.83$\times$ and 2.19$\times$ speedups for single and concurrent workflows~\cite{kvflow2025}. These results establish exact prefix reuse as a practical baseline, but its coverage remains limited.

\emph{Non-prefix reuse} extends the reuse unit to contiguous segments at arbitrary positions. CacheBlend selectively recomputed boundary tokens and reduced TTFT by 2.2--3.3$\times$ while improving throughput by 2.8--5$\times$ with negligible quality loss~\cite{cacheblend2025}. SparseX used sparse indexes to identify tokens that needed correction and supported interleaved reuse across dialogue, retrieval, and multi-agent workloads~\cite{sparsex2026}. RedKnot grouped caches by attention-head class and preserved the native layout of hybrid-attention models such as GDN. It reported 1.6--3.5$\times$ TTFT speedups and 4.7--7.8$\times$ higher concurrent-session throughput~\cite{redknot2026}.

\emph{Semantic-similarity reuse} links lexically different prompts that have similar meanings. SemShareKV used fuzzy embedding matching and positional correction, achieving a 6.25$\times$ speedup and 42\% memory savings on 5K-token summarization inputs~\cite{semsharekv2025}. Sim-LLM reported up to 39.40\% higher throughput and 34.65\% lower memory footprint on A40 and A100 GPUs~\cite{simllm2025}.

In \emph{multi-agent scenarios}, shared contexts often appear at different offsets. KVCOMM calibrated these offsets, reached reuse rates above 70\%, and reported up to 7.8$\times$ prefill speedup in five-agent workloads~\cite{kvcomm2025}. Agent Primitives transferred reusable review and planning contexts through KV caches, improving accuracy by 12.0--16.5\% and reducing token use and latency by about 3--4$\times$ relative to text-based coordination~\cite{agentprimitives2026}. RelayCaching reused the cache from one agent's decode phase in the next agent's prefill phase, achieving over 80\% reuse and up to 4.7$\times$ lower TTFT~\cite{relaycaching2026}. Taken together, these studies report up to 7.8$\times$ prefill speedup and over 80\% reuse, but broader coverage requires additional matching, calibration, and correction.

\subsection{KV Cache Compression}\label{sec:feasibility_compression}

KV cache compression balances size reduction, fidelity, and compression cost. Existing methods mainly use quantization, low-rank decomposition, or a combination of both.

\emph{Quantization methods} reduce bfloat16 (BF16) KV caches to lower bit widths. RotateKV reached 2-bit precision with a 3.97$\times$ peak-memory reduction and 2.32$\times$ decode speedup~\cite{rotatekv2025}. MPOQ reduced KV memory by about 75\% at 4-bit precision with nearly lossless quality~\cite{mpoq2025}. KVQuant supported up to 1M-token contexts on one A100-80GB and 10M-token contexts on eight GPUs~\cite{kvquant2024}. ZipCache combined salient-token selection with mixed precision and reported 4.98$\times$ compression with a 0.38\% accuracy drop~\cite{zipcache2024}. TurboQuant was quality-neutral at 3.5 bits per channel and had only slight degradation at 2.5 bits per channel~\cite{turboquant2026}.

\emph{Low-rank and mixed methods} exploit redundancy across hidden dimensions and layers. STAR-KV combined adaptive low-rank decomposition with quantization and reported up to 20$\times$ overall compression~\cite{starkv2026}. TailorKV selected layer-sensitive quantization and offloading policies, enabling a 128K context on one RTX 3090 at 82\,ms/token with near-lossless quality~\cite{tailorkv2025}. Kelle co-designed KV caching with eDRAM and reported 3.9$\times$ speedup and 4.5$\times$ higher energy efficiency on edge devices~\cite{kelle2025}. These results show that 4--20$\times$ size reduction is feasible, but the chosen ratio must still respect fidelity and computation constraints.

\subsection{Cross-Model KV Cache Mapping}\label{sec:feasibility_mapping}

Cross-model mapping transforms a source model's KV cache into a form usable by a target model. The main difficulty is the mismatch in layers, hidden dimensions, and attention structures. Existing work covers same-architecture reuse and cross-architecture mapping.

\emph{Same-architecture reuse} targets models with the same architecture but different parameters. DroidSpeak selectively recomputed important layers and reused the rest~\cite{droidspeak2025}. LRAgent separated base and adapter components to share caches across low-rank adaptation (LoRA) variants~\cite{lragent2026}. Activated LoRA extended this direction to serving-engine support for KV cache reuse across adapter switches~\cite{alora2025}. ICaRus enabled cache sharing across models through a logical encoder and decoder decomposition~\cite{icarus2026}.

\emph{Cross-architecture mapping} learns a transformation between incompatible KV cache spaces. C2C used neural projection and fusion to transfer a cache without per-token text communication and improved average accuracy by 8.5--10.5\% on policy-semantics tasks~\cite{c2c2026}. MoT used a mixture of translators for different architectures~\cite{mot2026}. For models in the same family, closed-form ridge regression offered a training-free alternative with much lower runtime than re-prefill~\cite{crossmodelmap2026}.

These methods operate inside serving engines or target specific model and adapter families. In contrast, the Unified AI Gateway treats reuse and mapping as candidate KV cache actions in a gateway-level decision space that also includes model selection, compute placement, cache placement, transfer, edge prefill, and target-side re-prefill.

For the gateway, mapping is useful when cache delivery and transformation together cost less than target-side recomputation while meeting the quality target. If $K_i(c)$ is the KV cache produced by source model $M_i$ for context $c$, switching to target model $M_j$ can use mapping when
\begin{equation}
T_{\mathrm{cache}}^{P_t\to e}(c)+T_{\mathrm{map}}^{i\to j}(c)<T_{\mathrm{recompute}}^{e,M_j}(c), \qquad \varepsilon_{i\to j}\leq\delta,
\label{eq:mapping_criterion}
\end{equation}
where $\varepsilon_{i\to j}$ is the mapping error and $\delta$ is the allowed quality loss. Depending on the task, $\varepsilon_{i\to j}$ can be measured as an accuracy drop, F1 loss, perplexity increase, or task-specific reward degradation relative to target-side re-prefill. The mapping cost grows with the transformed cache, whereas recomputation depends on context length, model size, and the attention implementation. Mapping is therefore often more attractive for long contexts. If the mapping criterion fails, the gateway uses direct transfer only when the source and target cache formats are compatible. Otherwise, it falls back to target-side re-prefill, as summarized in Table~\ref{tab:mapping} and Eq.~\eqref{eq:mapping_criterion}.

\begin{table*}[t!]
\centering
\caption{Conditional comparison of cross-model KV cache mapping and target-side re-prefill.}
\label{tab:mapping}
\rowcolors{2}{gray!8}{white}
\begin{tabular}{@{}>{\raggedright\arraybackslash}p{0.15\textwidth} 
                 >{\raggedright\arraybackslash}p{0.42\textwidth} 
                 >{\raggedright\arraybackslash}p{0.40\textwidth}@{}}
\toprule
Dimension & KV cache mapping & Re-prefill \\
\midrule
Compute object & Source KV cache (projection, reconstruction, or sparse patching) & Complete context (per-layer attention and feedforward recomputation) \\
Compute cost structure & Grows with the volume of the transformed cache and the mapping operator & Grows with context length and model size, with the exact scaling determined by the attention implementation \\
End-to-end latency & Decode starts once cache transformation completes; TTFT can be lower when mapping overhead is below re-prefill cost & Must wait for the complete forward pass. TTFT grows with context \\
Context preservation & Preserves source-context information subject to source/target alignment error & Recomputes the context in the target model's own KV cache space and avoids cross-model mapping error \\
Measured reference & Closed-form linear mapping 2.7--25$\times$ faster than re-prefill~\cite{crossmodelmap2026}; selective recomputation avoids full recomputation~\cite{droidspeak2025}; low-rank reconstruction and sparse patching achieve up to 2.65$\times$ TTFT speedup with quality loss no more than 5\%~\cite{scd2026} & Full recomputation is the main cost source of model switching (Section~\ref{sec:req_cache_mapping}) \\
\bottomrule
\end{tabular}
\end{table*}

These results show that cross-model mapping can run 2.7--25$\times$ faster than re-prefill and provide up to 2.65$\times$ TTFT speedup when the mapping satisfies the quality limit. The attainable gain still depends on model compatibility and mapping fidelity.

\subsection{Distributed KV Cache Storage and Transfer}\label{sec:feasibility_storage}

The storage layer must provide low-latency access under limited edge capacity. LMCache separated KV caches from GPU memory and supported sharing, lookup, eviction, migration, and compression across engines. Combined with vLLM, it reported up to 15$\times$ higher throughput and at least 2$\times$ lower latency on multi-turn workloads~\cite{lmcache2025}. ObjectCache delivered data in GPU consumption order and used object storage as an elastic tier. A 64K-token context added only 5.6\% TTFT over local DRAM on a 100\,Gbps RoCE cluster, while its bandwidth-aware scheduler further reduced TTFT by 1.2--1.8$\times$~\cite{objectcache2026}. Predictive Multi-Tier Memory Management expanded effective cache capacity from 40\,GB to over 38\,TB and reported 70--84\% hit rates in replay experiments~\cite{predictivemtm2026}. These results support multi-tier storage as a practical way to trade capacity against access latency.

Transfer optimization addresses bandwidth and RTT limits on wide-area links. Preble used shared prefix caching and distributed prompt scheduling to reduce average latency by 1.5--14.5$\times$ and p99 latency by 2--10$\times$~\cite{preble2025}. SmartGen reported that transferring a 48K-token cache over 25\,Gbps could take 6.5$\times$ the prefill time and 42.2\% of job completion time, while its selective push and pull policy reduced second-token latency by up to 4.3$\times$~\cite{smartgen2026}. SplitZip provided GPU-friendly lossless compression and improved end-to-end transfer by up to 1.32$\times$~\cite{splitzip2026}. KVServe selected compression online according to bandwidth and reported up to 9.13$\times$ lower job-completion time~\cite{kvserve2026}. Prefill-as-a-Service added a schedulable prefill tier across data centers and reported 54\% higher throughput and 64\% lower P90 TTFT in a heterogeneous deployment~\cite{prefillaservice2026}. FlowKV reduced average KV transfer latency by 96\% and accelerated inference by 15.2\%--48.9\% through optimized transfer, load-aware scheduling, and flexible PD-node allocation~\cite{flowkv2025}. KVDirect reduced per-request latency by 55\% through tensor-centric communication, pull-based KV transfer, and dynamic GPU scheduling~\cite{kvdirect2025}. Together, these storage and transfer systems report up to 15$\times$ higher throughput, at least 2$\times$ lower latency, and up to 96\% lower transfer latency. The realized gain depends on bandwidth, cache locality, and scheduling.

\subsection{Model Routing}\label{sec:feasibility_routing}

Model routing assigns requests to models under quality, latency, and cost constraints. OmniRouter reduced compute cost by at least 10.15\% and improved accuracy by up to 6.30\% through constrained optimization~\cite{omnirouter2025}. PORT approached the offline optimum without training and reported 4.25$\times$ higher throughput~\cite{port2025}. RouteLLM reduced invocation cost by more than 50\% while preserving quality, and FrugalGPT reported savings of up to 98\% through model cascades~\cite{routellm2025,frugalgpt2024}. Recent studies formulated KV cache constraints in online scheduling and jointly analyzed cache eviction and query routing~\cite{kvconstraints2025,kvrouting2026}. Robust KV Cache Management further optimized request routing, prefix caching, and GPU resource configuration under output-length uncertainty~\cite{robustkvcache2026}. Prior work also studied history-aware agent routing~\cite{strmac2025} and prefix-affinity routing in CacheRoute~\cite{cacheroute2026}. These approaches coordinate resources inside datacenter serving clusters. They report cost savings above 50\% and throughput gains up to 4.25$\times$, but do not measure the additional benefit of cross-model mapping and cross-tier placement at the gateway level.

\section{Workload-Level Benefit Analysis}\label{sec:benefit_analysis}

Section~\ref{sec:evidence} has established the component-level progress behind the Unified AI Gateway. This section examines how these KV cache capabilities combine under typical workload conditions.

\subsection{Workload Profiles and Settings}\label{sec:workload_setting}

We use an explicit workload-level KV cache preparation latency model to quantify the benefit of coordinated cache handling. Each workload is represented by $\theta_w=(L_{\mathrm{in}},L_{\mathrm{out}},h,r_{\mathrm{switch}},B)$, where $L_{\mathrm{in}}$ and $L_{\mathrm{out}}$ are the input and output lengths, $h$ is the effective KV cache hit rate, $r_{\mathrm{switch}}$ is the model-switching rate, and $B$ is the available bandwidth. We instantiate eight typical workload profiles. The five Preble profiles use its reported context, output, and shared-prefix characteristics~\cite{preble2025}. Multi-turn chat and retrieval-augmented generation (RAG) use representative hit rates for persistent dialogue and non-prefix document reuse~\cite{cachedattention2024,cacheblend2025}. The long-horizon coding-agent profile follows a vLLM-Mooncake trace with an approximately 80K-token turn-30 context, a 131:1 input-to-output ratio, and a 94.2\% hit rate. Its output length is rounded to 610 tokens~\cite{vllmmooncakestore2026}.

We evaluate switching rates of 25\%, 50\%, and 75\%. The 50\% setting approximates the 45\% per-turn switching rate reported for the single-turn routing baseline in vLLM's long-horizon agent evaluation, while 25\% and 75\% provide lower and stress-test settings~\cite{vllmsaar2026}. Here, the rate is the probability that a request with an available KV cache is routed to a different model. The reference bandwidth is 8~GB/s for the cross-domain setting. Table~\ref{tab:simulation_parameters} lists the remaining calibration settings, including the fixed 15~KB/token effective footprint and 4$\times$ compression ratio. These values isolate the effects of workload shape, cache availability, switching rate, and bandwidth. The footprint calibration reflects modern hybrid-attention models~\cite{prefillaservice2026,deepseekv42026,glm53flash2026}.

\begin{table}[t!]
\centering
\caption{Reference settings for the workload-level analytical simulation.}
\label{tab:simulation_parameters}
\scriptsize
\setlength{\tabcolsep}{3pt}
\rowcolors{2}{gray!10}{white}
\begin{tabular}{@{}p{0.31\columnwidth} p{0.20\columnwidth} p{0.39\columnwidth}@{}}
\toprule
Parameter & Setting & Basis or role \\
\midrule
KV cache footprint $s_{\mathrm{KV}}$ & 15~KB/token & Representative effective footprint for modern hybrid-attention models~\cite{prefillaservice2026,deepseekv42026,glm53flash2026} \\
Compression ratio $\rho$ & 4$\times$ & Conservative reference near reported 4--5$\times$ reductions~\cite{rotatekv2025,zipcache2024} \\
Prefill coefficients $a_{\mathrm{prefill}}$ / $b_{\mathrm{attn}}$ / $\tau_{\mathrm{prefill}}$ & 0.04~ms/token / $3\times10^{-6}$~ms/token$^2$ / 20~ms & Rounded representative values based on the A100 profiles released with DistServe~\cite{distserve2024,distserveartifact2024} \\
Decode coefficient $t_{\mathrm{decode}}$ & 10~ms/token & Reference setting for the secondary end-to-end diagnostic \\
Routing $t_{\mathrm{route}}$ / lookup $t_{\mathrm{lookup}}$ budget & 2 / 3~ms & Control-plane calibration. Routing scale informed by prior work~\cite{prefilldeflection2026} \\
Model-switching rate $r_{\mathrm{switch}}$ & 25\% / 50\% / 75\% & Sensitivity settings for the probability of switching away from the cache-producing model \\
Cached / uncached input price $\eta=c_{\mathrm{hit}}/c_{\mathrm{miss}}$ & 0.10 & Representative cached-input discount~\cite{anthropicpricing2026} \\
Output / uncached input price $\lambda=c_{\mathrm{out}}/c_{\mathrm{miss}}$ & 5.0 & Representative output-to-input pricing relationship~\cite{anthropicpricing2026} \\
Available bandwidth $B$ & 8~GB/s & Maximum available bandwidth in the reference cross-domain setting~\cite{smartgen2026,objectcache2026} \\
\bottomrule
\end{tabular}
\end{table}

\subsection{KV-Cache Model and Metrics}\label{sec:benefit_metrics}

The scalar model below instantiates the latency component of the general per-request quantities in Section~\ref{sec:joint_decision} for a reference model pool and workload. In the general formulation, $C_{tjea}$ denotes gateway-side compute, memory, storage, and network cost, whereas this workload-level evaluation focuses on user-visible latency and uses $T$ for latency-equivalent preparation times. Here $h$, $B$, and the stated feasibility conditions summarize the current KV cache environment, and $h$ is the effective cache-hit rate after background placement and replication.

We assume that background placement, replication, eviction, and migration have already prepared this cache state, so their costs are amortized over the request stream and omitted from the per-request cache-preparation critical path. The results therefore condition on effective state management rather than evaluating its policy or overhead.

For a workload row, let $L=L_{\mathrm{in}}$. The cache-preparation action times are given in Eq.~\eqref{eq:workload_action_costs}, shown as follows
\begin{equation}
\begin{aligned}
T_{\mathrm{prefill}}(L)&=a_{\mathrm{prefill}}L+b_{\mathrm{attn}}L^2+\tau_{\mathrm{prefill}},\\
 S(L)&=L s_{\mathrm{KV}},\\
T_{\mathrm{transfer}}&=\frac{S(L)}{\rho B},\\
T_{\mathrm{re-prefill}}&=T_{\mathrm{prefill}}(L),
\end{aligned}
\label{eq:workload_action_costs}
\end{equation}
where $a_{\mathrm{prefill}}$, $b_{\mathrm{attn}}$, and $\tau_{\mathrm{prefill}}$ are the linear projection coefficient, quadratic full-attention coefficient, and fixed prefill overhead listed in Table~\ref{tab:simulation_parameters}. The linear-plus-quadratic form follows DistServe. We use rounded representative coefficients based on the A100 profiles released with its artifact for workload-level comparison~\cite{distserve2024,distserveartifact2024}. The KV cache size $S(L)$ remains linear in context length. Here, $\rho=4$ is the compression ratio.

Recent measurements show closed-form cross-model mappers running 2.7--25$\times$ faster than re-prefill, while multi-tier cache systems demonstrate overlap between I/O and cache processing~\cite{crossmodelmap2026,kvdrive2026}. Motivated by these results, the reference model pipelines mapping with chunked cache transfer and places compression, decompression, and memory-copy operations outside the critical path. RTT is not modeled separately because it is common to the network path. Under this overlap assumption, the numerical model sets $T_{\mathrm{map+transfer}}=T_{\mathrm{transfer}}$.

A same-model hit reuses the cache directly, a switching hit uses mapping and transfer, and a miss triggers target-side full prefill. Thus, the switching path compares $T_{\mathrm{map+transfer}}$ with $T_{\mathrm{re-prefill}}$, while the input-token cost effect is calculated separately below. Therefore, the Unified AI Gateway preparation time is
\begin{equation}
\begin{aligned}
T_{\mathrm{prep}}^{\mathrm{UG}}&=t_{\mathrm{route}}+t_{\mathrm{lookup}}+h r_{\mathrm{switch}}T_{\mathrm{switch}}\\
&\quad+(1-h)T_{\mathrm{re-prefill}},\\
T_{\mathrm{switch}}&=\min\{T_{\mathrm{map+transfer}},T_{\mathrm{re-prefill}}\},\\
T_{\mathrm{prep}}^{\mathrm{base}}&=t_{\mathrm{route}}+t_{\mathrm{lookup}}\\
&\quad+[(1-h)+h r_{\mathrm{switch}}]T_{\mathrm{re-prefill}}.
\end{aligned}
\label{eq:workload_gateway_time}
\end{equation}
Every request incurs one cache-lookup time, regardless of whether the lookup succeeds. A successful lookup avoids the subsequent cache-preparation time. Equation~\eqref{eq:workload_gateway_time} gives the resulting Unified AI Gateway and baseline preparation times. The baseline represents a conventional AI gateway that selects a target model from a model pool. A compatible same-model KV cache may be reused within the serving runtime, while a cache hit followed by a model switch and a cache miss both require target-side full prefill. Define $T_{\mathrm{e2e}}^{\mathrm{base}}=T_{\mathrm{prep}}^{\mathrm{base}}+L_{\mathrm{out}}t_{\mathrm{decode}}$ and $T_{\mathrm{e2e}}^{\mathrm{UG}}=T_{\mathrm{prep}}^{\mathrm{UG}}+L_{\mathrm{out}}t_{\mathrm{decode}}$. The primary latency metric is the preparation speedup, which serves as a TTFT proxy in this model. We retain end-to-end speedup as a secondary diagnostic for the effect of output generation. The two speedup metrics are
\begin{equation}
\begin{aligned}
S_{\mathrm{e2e}}(\theta_w)&=\frac{T_{\mathrm{e2e}}^{\mathrm{base}}}{T_{\mathrm{e2e}}^{\mathrm{UG}}}, &
S_{\mathrm{prep}}(\theta_w)&=\frac{T_{\mathrm{prep}}^{\mathrm{base}}}{T_{\mathrm{prep}}^{\mathrm{UG}}}.
\end{aligned}
\label{eq:workload_speedup}
\end{equation}
The end-to-end ratio is retained as a whole-request diagnostic. The preparation speedup in Eq.~\eqref{eq:workload_speedup} is reported as the primary TTFT result in Table~\ref{tab:benefits} and Figure~\ref{fig:scenario_benefits}.

We calculate the input-token cost ratio separately from latency. Let $c_{\mathrm{miss}}$ and $c_{\mathrm{hit}}$ denote the prices of uncached and cached input tokens, respectively, and let $\eta=c_{\mathrm{hit}}/c_{\mathrm{miss}}$. Current commercial model services provide a representative reference. OpenAI's current token-based rate card and Anthropic's cache-read rate set cached input at 0.10 times the corresponding uncached input price for listed models, equivalent to a 10$\times$ price reduction~\cite{openaipricing2026,anthropicpricing2026}. We therefore use $\eta=0.10$. Anthropic's current API schedule lists output tokens at five times the uncached input-token price for Sonnet 4.6, so we use $\lambda=c_{\mathrm{out}}/c_{\mathrm{miss}}=5.0$ for the end-to-end token-cost calculation~\cite{anthropicpricing2026}.

Under the reference setting, every cache hit, including a hit followed by model switching, is assumed to be mapped successfully. Mapping is treated as a lightweight or linear transformation whose computation is hidden by overlap~\cite{crossmodelmap2026,kvdrive2026}. Same-model hits and mapped switching hits therefore use the cached-input price, while cache misses use the uncached price. The normalized costs and their ratio are
\begin{equation}
\begin{aligned}
\frac{C_{\mathrm{input}}^{\mathrm{base}}}{c_{\mathrm{miss}}}
&=h(1-r_{\mathrm{switch}})\eta
 +h r_{\mathrm{switch}}+(1-h),\\
\frac{C_{\mathrm{input}}^{\mathrm{UG}}}{c_{\mathrm{miss}}}
&=h\eta+(1-h),\\
G_{\mathrm{input}}(\theta_w)&=
\frac{C_{\mathrm{input}}^{\mathrm{base}}}{C_{\mathrm{input}}^{\mathrm{UG}}}
 =\frac{h(1-r_{\mathrm{switch}})\eta
 +h r_{\mathrm{switch}}+(1-h)}
  {h\eta+(1-h)}.
\end{aligned}
\label{eq:input_cost_ratio}
\end{equation}
This calculation complements the resource-cost term $C_{tjea}$ in the Global Control Plane formulation with an input-billing metric. It preserves cached-input billing for cache hits, including successful cross-model mapping, while ordinary misses and output-token charges remain unchanged. To include output-token charges, let $\bar C_{\mathrm{base}}=C_{\mathrm{input}}^{\mathrm{base}}/c_{\mathrm{miss}}$ and $\bar C_{\mathrm{UG}}=C_{\mathrm{input}}^{\mathrm{UG}}/c_{\mathrm{miss}}$. The end-to-end token-cost ratio is
\begin{equation}
G_{\mathrm{token}}^{\mathrm{e2e}}(\theta_w)=
\frac{L_{\mathrm{in}}\bar C_{\mathrm{base}}+\lambda L_{\mathrm{out}}}
{L_{\mathrm{in}}\bar C_{\mathrm{UG}}+\lambda L_{\mathrm{out}}}.
\label{eq:e2e_token_cost_ratio}
\end{equation}
The input-cost and end-to-end (E2E) cost columns in Table~\ref{tab:benefits} are computed with Eqs.~\eqref{eq:input_cost_ratio} and~\eqref{eq:e2e_token_cost_ratio}, respectively.

\subsection{Simulation Results and Sensitivity}\label{sec:simulation_results}

Table~\ref{tab:benefits} compares the modeled TTFT, input-cost, and E2E cost benefits of the eight workload profiles at model-switching probabilities of 25\%, 50\%, and 75\%.

\begin{table*}[t!]
\centering
\caption{Workload-level settings and simulated benefits for typical Unified AI Gateway workloads. Context and output lengths are mean values for the Preble workloads, while the multi-turn chat and RAG settings represent persistent dialogue and non-prefix document reuse reported in prior work. Reported shared-prefix proportions serve as effective cache-hit rates for the Preble profiles. The three benefit columns list results for $r_{\mathrm{switch}}=25\%$, 50\%, and 75\%, in that order. The input-cost ratio uses a 10$\times$ cached-input discount, while the E2E cost ratio also includes output-token charges with $\lambda=5.0$.}
\label{tab:benefits}
\scriptsize
\setlength{\tabcolsep}{1.5pt}
\renewcommand{\arraystretch}{1.05}
\rowcolors{2}{gray!8}{white}
\begin{tabular}{@{}>{\raggedright\arraybackslash}p{0.14\textwidth}
                >{\centering\arraybackslash}p{0.075\textwidth}
                >{\raggedright\arraybackslash}p{0.135\textwidth}
                >{\centering\arraybackslash}p{0.055\textwidth}
                >{\centering\arraybackslash}p{0.185\textwidth}
                >{\centering\arraybackslash}p{0.185\textwidth}
                >{\centering\arraybackslash}p{0.185\textwidth}@{}}
\toprule
Workload & \makecell{Context/Output\\$L_{\mathrm{in}}/L_{\mathrm{out}}$} & KV cache reuse pattern & \makecell{Cache\\hit rate $h$} & \makecell{TTFT speedup\\25\% / 50\% / 75\%} & \makecell{Input-cost ratio\\25\% / 50\% / 75\%} & \makecell{E2E cost ratio\\25\% / 50\% / 75\%} \\
\midrule
Multi-turn chat~\cite{cachedattention2024} & 16K/200 & Historical dialogue & 0.70 & 1.57$\times$ / 2.14$\times$ / 2.71$\times$ & 1.43$\times$ / 1.85$\times$ / 2.28$\times$ & 1.36$\times$ / 1.73$\times$ / 2.09$\times$ \\
RAG~\cite{cacheblend2025} & 16K/300 & Non-prefix chunks & 0.50 & 1.25$\times$ / 1.49$\times$ / 1.74$\times$ & 1.20$\times$ / 1.41$\times$ / 1.61$\times$ & 1.17$\times$ / 1.35$\times$ / 1.52$\times$ \\
ToolBench~\cite{preble2025,infercept2024} & 1.8K/43 & Shared tool instructions & 0.85 & 2.05$\times$ / 3.09$\times$ / 4.11$\times$ & 1.81$\times$ / 2.63$\times$ / 3.44$\times$ & 1.54$\times$ / 2.09$\times$ / 2.63$\times$ \\
Embodied agent~\cite{preble2025,continuum2025} & 2.3K/16 & Shared context and pauses & 0.97 & 4.37$\times$ / 7.55$\times$ / 10.56$\times$ & 2.72$\times$ / 4.44$\times$ / 6.16$\times$ & 2.35$\times$ / 3.69$\times$ / 5.04$\times$ \\
Programming~\cite{preble2025,cachewise2026} & 3.9K/190 & Shared prompts and context & 0.97 & 5.39$\times$ / 9.48$\times$ / 13.28$\times$ & 2.72$\times$ / 4.44$\times$ / 6.16$\times$ & 1.59$\times$ / 2.17$\times$ / 2.76$\times$ \\
Video QA~\cite{preble2025} & 9.9K/4 & Repeated video context & 0.88 & 2.70$\times$ / 4.36$\times$ / 5.99$\times$ & 1.95$\times$ / 2.90$\times$ / 3.86$\times$ & 1.94$\times$ / 2.89$\times$ / 3.83$\times$ \\
Long-document QA (LooGLE)~\cite{preble2025,scbench2025} & 23.5K/16 & Shared long-document context & 0.91 & 3.44$\times$ / 5.83$\times$ / 8.17$\times$ & 2.13$\times$ / 3.26$\times$ / 4.39$\times$ & 2.11$\times$ / 3.22$\times$ / 4.33$\times$ \\
Long-horizon coding agent~\cite{vllmmooncakestore2026} & 80K/610 & Persistent cross-turn prefix & 0.942 & 5.01$\times$ / 8.97$\times$ / 12.87$\times$ & 2.39$\times$ / 3.79$\times$ / 5.18$\times$ & 2.11$\times$ / 3.23$\times$ / 4.34$\times$ \\
\bottomrule
\end{tabular}
\end{table*}

To isolate the benefit of coordinated cache handling, we hold the conventional AI gateway's model-selection policy fixed and treat its model-switching probability as the exogenous parameter $r_{\mathrm{switch}}$. We then apply the Unified AI Gateway cache actions to the same workload profiles. This comparison measures cache-handling gains at the incumbent switching rate rather than evaluating a learned or optimized joint routing policy. Because prepared or mapped caches reduce the cost of changing models, a joint policy could create more switching opportunities. The 50\% and 75\% settings quantify the additional benefit available as the effective switching rate increases under the stated lightweight-mapping and overlap assumptions.

The results depend on workload shape and switching rate. Across the 25\%, 50\%, and 75\% switching settings, the eight profiles achieve TTFT speedups of 1.25$\times$--5.39$\times$, 1.49$\times$--9.48$\times$, and 1.74$\times$--13.28$\times$, respectively. The corresponding input-cost ratios are 1.20$\times$--2.72$\times$, 1.41$\times$--4.44$\times$, and 1.61$\times$--6.16$\times$, while the E2E cost ratios are 1.17$\times$--2.35$\times$, 1.35$\times$--3.69$\times$, and 1.52$\times$--5.04$\times$. For the long-horizon coding-agent profile, TTFT speedup grows from 5.01$\times$ to 12.87$\times$, the input-cost ratio from 2.39$\times$ to 5.18$\times$, and the E2E cost ratio from 2.11$\times$ to 4.34$\times$ as the switching rate increases from 25\% to 75\%.

High-hit, long-context workloads benefit most because each successful cache action avoids more prefill work and a larger uncached-input charge. This effect is especially strong for agentic workloads with long inputs and short outputs~\cite{preble2025,continuum2025,vllmmooncakestore2026}. During external pauses, tool-augmented and coding agents also require cache retention, offloading, eviction, and reload~\cite{infercept2024,continuum2025,cachewise2026}. Within the modeled setting, successful mapping and overlapped mapping computation make the TTFT and cost benefits grow with the switching rate. Lower cache-aware switching costs could also let a joint routing policy choose lower-cost or better-suited models more often, but this effect is outside the fixed-policy comparison.

\begin{figure*}[t!]
\centering
\begin{subfigure}[t]{0.49\textwidth}
    \centering
    \includegraphics[width=\linewidth]{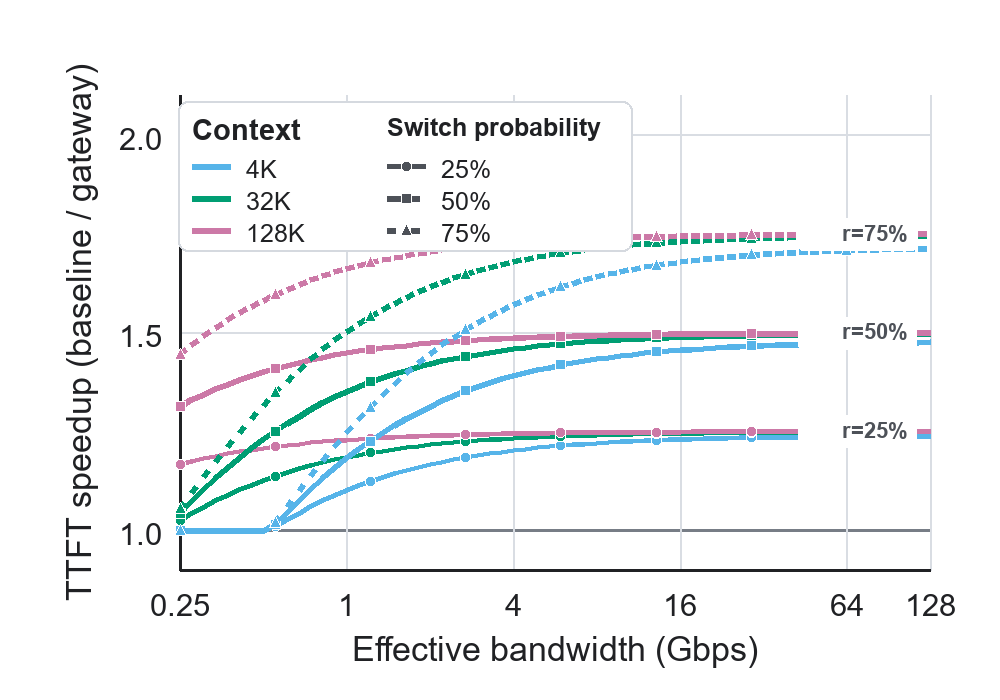}
    \caption{TTFT speedup versus bandwidth.}
    \label{fig:scenario_bandwidth}
\end{subfigure}
\hfill
\begin{subfigure}[t]{0.49\textwidth}
    \centering
    \includegraphics[width=\linewidth]{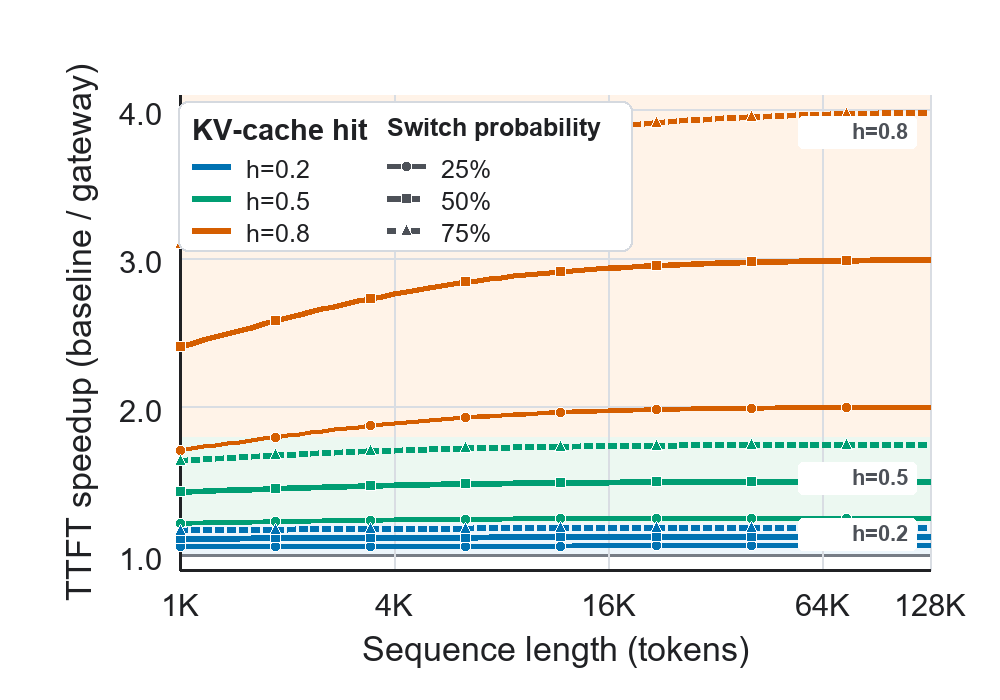}
    \caption{TTFT speedup versus sequence length.}
    \label{fig:scenario_sequence}
\end{subfigure}
\caption{TTFT speedup under variations in bandwidth, sequence length, cache-hit rate, and model-switching rate. Values are normalized to the conventional-gateway baseline defined in Eq.~\eqref{eq:workload_gateway_time}, and the horizontal line at 1.0 denotes equal latency. Panel (a) uses $h=0.5$, while Panel (b) uses 64~Gbps. Colors encode context length in Panel (a) and cache-hit rate in Panel (b), and solid, dashed, and dotted lines represent switching rates of 25\%, 50\%, and 75\%, respectively. The light blue, green, and orange background bands in Panel (b) correspond to the curve groups with cache-hit rates of $h=0.2$, $h=0.5$, and $h=0.8$, respectively.}
\label{fig:scenario_benefits}
\end{figure*}

Figure~\ref{fig:scenario_benefits} shows that higher cache availability and switching rates produce larger ratios, while bandwidth and context length determine the attainable range. In deployment, the Unified AI Gateway can use lower switching costs to increase the model-switching probability and coordinate routing and cache actions at higher-benefit operating points. The resulting higher switching probability may provide additional gains beyond the fixed-policy results reported here.

\section{Deployment Readiness and Research Directions}\label{sec:limitations}

Section~\ref{sec:evidence} summarizes the current implementation status of the main gateway components, while Section~\ref{sec:benefit_analysis} analyzes their workload-level benefits through analytical simulation. Realizing these benefits in practice still requires addressing several gaps in current systems. This section assesses deployment readiness and identifies the remaining research directions.

\subsection{Cross-Model Semantic Alignment}\label{sec:challenge_alignment}

Cross-model KV cache mapping must preserve the semantics encoded by high-dimensional caches across heterogeneous models, since layer-wise errors can accumulate through attention and feedforward blocks. C2C and MoT demonstrated neural projection and mixture-of-translators on selected model pairs, but required pair-specific training~\cite{c2c2026,mot2026}. SCD, dense alignment, and sparse mapping reduced this burden in selected settings, but they still involved quality, training, or model-family constraints~\cite{scd2026,denselatent2026,crossmodelmap2026}. As the model pool grows, the gateway therefore needs mapping error bounds, low-sample or training-free adaptation, and benchmarks across models, tasks, and languages.

\subsection{Heterogeneous Edge Hardware}\label{sec:challenge_hardware}

Edge nodes differ in compute units, memory hierarchies, interconnects, and power limits, so GPU-oriented operators and layouts cannot be transferred directly to neural processing unit (NPU), field-programmable gate array (FPGA), or application-specific integrated circuit (ASIC) platforms~\cite{hardwareaccel2025}. Datacenter KV methods also assume high-bandwidth memory (HBM) and large batches, whereas edge devices make cache residence, eviction, and reload more costly. SwiftCache and Kelle showed the value of platform-specific cache and memory co-design~\cite{swiftcache2026,kelle2025}. ENEC further illustrated this hardware dependence with lossless model-weight compression on Ascend NPUs~\cite{enec2026}. A gateway-wide solution must adapt compression, placement, numerical format, and scheduling to device memory, power, and thermal profiles.

\subsection{Distributed KV Cache Consistency}\label{sec:challenge_consistency}

Distributed placement makes KV cache consistency a performance concern. Because a KV cache can be recomputed from the context, stale metadata usually causes a miss and repeated prefill rather than an incorrect result, which motivated asynchronous refresh and weak consistency in P2P inference~\cite{p2pinference2026}. TraCT, SwiftCache, and Prefill-as-a-Service illustrated different approaches to shared access, ownership reassignment, and separated cache generation and consumption~\cite{tract2025,swiftcache2026,prefillaservice2026}. The gateway must co-design consistency, mobility, placement, and routing so that migration and metadata maintenance do not erase reuse gains.

\subsection{Security and Privacy}\label{sec:challenge_security}

KV caches encode user prompts and generated content, so edge storage and sharing create reconstruction and timing-channel risks. PROMPTPEEK, black-box analyses, SpliceLeak, and KV-Cloak show that these risks apply to prefix hits, non-prefix fusion, and direct tensor access~\cite{promptpeek2025,earlybird2025,spliceleak2026,shadowinthecache2026}. PrefixWall shows that selective isolation can retain part of the reuse benefit~\cite{prefixwall2026}. Unified gateway control therefore requires authorization, separation of users' data, auditing, timing-signal protection, encryption, and data-residency enforcement.

\subsection{System Economics}\label{sec:challenge_cost}

Caching, compression, mapping, and migration change compute, storage, and network costs simultaneously, while prices and bottlenecks vary across locations and time. Existing analyses cover parts of this trade-off, but do not connect hit rate, context length, cache lifecycle, bandwidth price, and service quality in one gateway objective~\cite{inferenceeconomics2025,beyondcontextwindow2026}. Inter-region traffic must therefore justify its latency benefit, and future models should include storage lifecycle, bandwidth billing, energy, hardware depreciation, and break-even conditions~\cite{prefillaservice2026,computingpowernetwork2022}.

\subsection{Standards and Ecosystem Support}\label{sec:challenge_standardization}

KV cache interoperability remains less mature than unified APIs. Engines differ in page size, cache layout, precision, and device affinity, making point-to-point connectors costly as model and engine versions grow. LMCache and emerging interoperation proposals provided common control events and cache descriptions, but compression, mapping, and transfer interfaces remained open~\cite{lmcache2025}. SCBench covered several cache operations but not cross-model mapping, cross-node transfer, or consistency semantics~\cite{scbench2025}. Progress therefore requires a common KV cache data model, lifecycle and transfer events, reference implementations, and cross-vendor benchmarks.

\subsection{KV Cache Transfer and Memory Management}\label{sec:coupling}

Transfer efficiency depends on memory layout and compression. PagedAttention's non-contiguous blocks can reduce PCIe efficiency during direct transfer, while compaction adds copies during decode~\cite{pagedattention2023}. SwiftCache, KVServe, CacheGen, and TraCT showed that reload, bandwidth-aware compression, streaming, and shared memory must be considered together~\cite{swiftcache2026,kvserve2026,cachegen2024,tract2025}. The gateway therefore needs page-aware transfer primitives and SLO-aware schedulers that jointly select reuse, compression, priority, and placement.

\subsection{KV Cache Portability and Architectural Evolution}\label{sec:challenge_transportability}

Model design and KV cache movement are still optimized separately. Surveys and recent systems expose a gap between mature token and system optimizations and less developed model-level portability objectives~\cite{kvsurvey2025,starkv2026,kvserve2026,scd2026}. Future models could incorporate low-rank cache structure, sparse attention, or compression regularization to make cache movement more efficient.

Although this paper focuses on KV caches, future gateway designs may extend the same coordination principle to other transferable inference representations used by recurrent, non-autoregressive, or diffusion architectures~\cite{mamba2023,rwkv2023,nonauto2018,diffusionlm2022}.

\section{Conclusion}\label{sec:conclusion}

This paper has defined the Unified AI Gateway setting, formalized its joint model-routing and KV cache decision, and connected the End--Edge--Cloud architecture to the responsibilities and requirements of its main components. It has also synthesized component-level evidence and analyzed the performance, input-cost, and model-switching benefits of coordinated KV cache handling. Across eight typical workload profiles and model-switching rates from 25\% to 75\%, the workload-level analytical simulation reports TTFT speedups of 1.25$\times$--13.28$\times$ and input-cost benefits of 1.20$\times$--6.16$\times$. The sensitivity analysis shows that these modeled benefits grow with reusable-cache coverage and switching frequency, provided that KV cache transfer can avoid target-side full prefill. Realizing this setting remains an end-to-end systems integration problem involving cache fidelity, distributed resources, and trustworthy coordination.

\bibliographystyle{IEEEtran}
\bibliography{bibliography}

\end{document}